\documentclass[letterpaper,useAMS,usenatbib]{mn2e}
 \usepackage{txfonts}
 \usepackage{graphicx}
 \title[A broadened O VIII line in the UCXB 4U 0614+091]
{A relativistically broadened O VIII Ly$\alpha$ line in the ultra-compact X-ray binary 4U 0614+091} 

\author [Madej et al.]
{O.K. Madej,$^{1,2}$\thanks{E-mail: O.Madej@sron.nl} P.G. Jonker,$^{2,3,4}$ A.C. Fabian,$^{5}$ C. Pinto,$^{2,1}$ F.  Verbunt,$^{1}$ J. de Plaa$^{2}$ \\
\\
\normalsize{$^{1}$Astronomical Institute, Utrecht University, PO Box 80000, 3508 TA Utrecht, The Netherlands}\\
\normalsize{$^{2}$SRON Netherlands Institute for Space Research, Sorbonnelaan 2, 3584 CA Utrecht,The Netherlands}\\
\normalsize{$^{3}$Harvard-Smithsonian Center for Astrophysics, 60 Garden Street, Cambridge, MA 02138, USA}\\
\normalsize{$^{4}$Radboud University Nijmegen, P.O. Box 9102, 6500 HC Nijmegen, The Netherlands}\\
\normalsize{$^{5}$Institute of Astronomy, Madingley Road, Cambridge CB3 OHA, UK} }

\begin{document}

\date{}

\maketitle
\label{firstpage}
\def\apjl{ApJ}
\def\aj{AJ}
\def\apj{ApJ}
\def\pasp{PASP}
\def\spie{SPIE}
\def\apjs{ApJS}
\def\araa{ARAA}
\def\aap{A\&A}
\def\nat{Nature}
\def\mnras{MNRAS}
\def\prd{Phys.Rev.D}
\begin{abstract}
Ultra-compact X-ray binaries consist of a neutron star or black hole that accretes material from a white dwarf-donor star. The ultra-compact nature is expressed in very short orbital periods of less than 1 hour. In the case of 4U~0614+091 oxygen-rich material from a CO or ONe white dwarf is flowing to the neutron star. This oxygen-rich disc can reflect X-rays emitted by the neutron star giving a characteristic emission spectrum. We have analyzed high-resolution RGS and broad band EPIC spectra of 4U 0614+091 obtained by the \textit{XMM-Newton} satellite. We detect a broad emission feature at $\sim0.7$ keV in both instruments, which cannot be explained by unusual abundances of oxygen and neon in the line of sight, as proposed before in the literature. We interpret this feature as O VIII Ly$\alpha$ emission caused by reflection of X-rays off highly ionized oxygen, in the strong gravitational field close to the neutron star. 

\end{abstract}


\section{Introduction}
The object of our study is the ultra-compact X-ray binary (UCXB) 4U 0614+091 consisting of a neutron star that accretes gas from a white dwarf. The source is placed close to the Galactic plane at a distance of 3.2 kpc \citep{Kuulk} towards the Galactic anticenter. The characteristic that distinguishes UCXBs from low-mass X-ray binaries (LMXBs) is a very short orbital period, it is typically less than 80 minutes. For this source, the measurements revealed an orbital period of $\sim50$ min \citep{Shahbaz}. The light curve of this object does not show signs of eclipses or dips, which indicates the upper limit of the inclination of the system to be around $70^{\circ}$. The source experiences type I X-ray bursts \citep{swank,brandt} and is classified as an atoll source \citep{mendez}, which means that it shows transitions between the low/hard (island state) and high/soft state (banana state).  \\
An interesting process occurring in neutron star LMXBs is reflection. The X-ray photons emitted from a neutron star can be reflected by the accretion disc, leading to an X-ray emission line spectrum and free-free continuum. If the infalling photons are reflected by the innermost part of the disc, where a strong gravitational field is present, the observed spectral features will be broadened by the relativistic Doppler effect and gravitational redshift \citep{fabian89}. The emission lines can be additionally affected by Compton scattering in the ionized disc material, which upscatters the line photons preferentially to higher energies \citep{ballant}. The most prominent reflection line in the case of a disc consisting of gas with solar abundances is that of Fe K$\alpha$ at $\sim6.4$ keV and second strongest is O VIII Ly$\alpha$ at $\sim0.65$ keV. So far only the Fe K$\alpha$ line has been reported in the LMXBs \citep{reis,cackett1,miller}. \\
\begin{figure*}
\vspace{-3mm}
\includegraphics[width=0.82\textwidth]{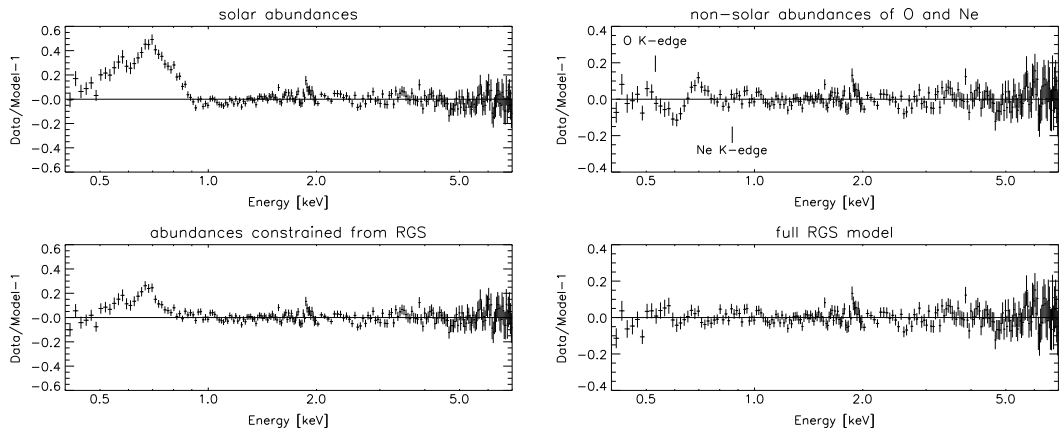}
  \caption{Residuals obtained by fitting an absorbed power-law model to the MOS2 data. Top-left: Model assuming an interstellar medium (ISM) with proto-solar abundances of \citet{lodders}. The region $0.6-0.8$ keV was excluded from the fit, Top-right: Model with an ISM where the abundance of oxygen and neon are allowed to vary. A similar model was proposed by \citet{juett} to explain the excess emission around 0.7 keV. The fit gives an overabundance of neon and an underabundance of oxygen. Bottom-left: Model with abundances of oxygen, iron and neon constrained from the RGS high-resolution spectra. The region $0.6-0.8$ keV was excluded from the fit. Bottom-right: Model with parameters of the Gaussian line obtained from the RGS spectra.}
\vspace{-5mm}
\end{figure*}
Evidence for reflection in 4U 0614+091 was previously claimed in BeppoSAX data \citep{piraino}. The authors interpreted the spectral continuum-like feature around $\sim60$ keV as the reflection from the optically-thick disc. A line-like feature at 0.7 keV  indicating a possible broad O VIII emission line or a set of O VII-O VIII and Fe XVII-Fe XIX lines was reported in the low-resolution \textit{EXOSAT} and \textit{ASCA} spectra \citep{christian,juett}. However, \citet{paerels} found no evidence for emission lines in the high-resolution \textit{Chandra} data. \citet{paerels} did report an overabundance of neon. \citet{juett} showed that the feature at 0.7 keV could be eliminated in the \textit{ASCA} spectra when the overabundance of neon is considered. However, the neutral absorption edges of oxygen, iron and neon are unresolved in those low-resolution spectra. Since the 0.7 keV feature appears close to these edges, it is crucial for the determination of the continuum level to measure their depths accurately. \\
We analyze simultaneous low and high-resolution spectra of 4U 0614+091 obtained by the \textit{XMM-Newton} satellite. The high-resolution spectra contain well resolved absorption edges, which allow us to improve the measurement of the abundances of oxygen, iron and neon with respect to the previous measurements done by \citet{juett}. The residuals show that besides the higher than solar abundance of neon there is a broad emission feature present in the data with a peak at the position of the O VIII Ly$\alpha$ line. 
\section{Observations and data reduction}
\textit{XMM-Newton} observed 4U 0614+091 on 2001 March 13 starting at 9:35 UT for $\sim$ 11 ksec  and immediately afterwards at 12:32 UT for $\sim$ 17 ksec. In both parts of the observation two Reflection Grating Spectrometers (RGS) were collecting data. RGS is a high-resolution spectrometer covering the energy range from 7 to 38~\AA\ ($0.3-2.1$ keV) with spectral resolution of $E/\Delta E=$100 to 500 \citep{denHerder}. In the second part of the observation the European Photon Imaging Cameras: MOS1, MOS2 and pn were collecting data simultaneously with RGS. MOS1, MOS2 and pn provide broad band spectral coverage with modest resolution ($E/\Delta E=$20 to 50) over the energy range of 0.3 to 10 keV \citep{turner}. The MOS1 and pn camera were operated in the Timing mode, while MOS2 was operated in the Full Frame (Imaging) mode.\\
We reduce the data using the \textit{XMM-Newton} Data Analysis software SAS version 9.0. The light curve shows no significant contamination from soft protons.
We extract source photons with pixel pattern equal to 0 from the MOS1 camera with {\sc rawx} from 300 to 320 and background with {\sc rawx} from 260 to 280 (beyond the source PSF). The net source count rate is $58.54\pm0.06$ c/s (full bandpass). We extract the MOS2 observation with pixel pattern below or equal to 12. The MOS2 observation suffers from pile-up. In order to reduce the effect of pile-up on the source spectrum, we exclude events from within the circle with a radius of 24 arcsec centered on the source position. The background spectrum for the MOS2 observation is extracted from CCD-3. The net source count rate is $13.01\pm0.03$ c/s. We extract source photons with pixel pattern less than 5 from the pn camera with  {\sc rawx} from 33 to 45 and background with {\sc rawx} from 15 to 25. The net source count rate is $238.74\pm0.14$ c/s. \\
The data collected by RGS are reduced using standard software pipeline which generates source and background spectra as well as response files. We fit the data using the {\sc spex} package\footnote[1]{http://www.sron.nl/spex} \citep{kaastra1996}. Errors on the fit parameters reported throughout this Manuscript correspond to a 68\% confidence level for each calculated parameter ($\Delta\chi^2=1$).
\section{Analysis and results}
\subsection{Model with high neon abundance}
\subsubsection{Broad band spectra}
The typical continuum model for X-ray binaries contains two components: a power-law, which is important in the island state and a black body, which is important in the banana state of the source. In our case adding a soft thermal component (black body) does not improve the fit, therefore we use only a power-law to fit the data. During the observation the source was in the island state \citep{mendez2002}, hence it is not surprising that the soft thermal component is not present. The X-ray spectrum of the source is affected by the interstellar medium. Most of the gas in the line of sight is neutral \citep{ferriere}, therefore as to a first approximation we use only a neutral absorption model ({\sc hot} model with very low temperature-neutral gas limit). This model calculates the transmission of a plasma in collisional ionization equilibrium. For the reference abundances of elements in the neutral gas we choose the proto-solar abundances of \citet{lodders}. 
\begin{figure*}  
\vspace{-3mm}
\includegraphics[width=0.75\textwidth]{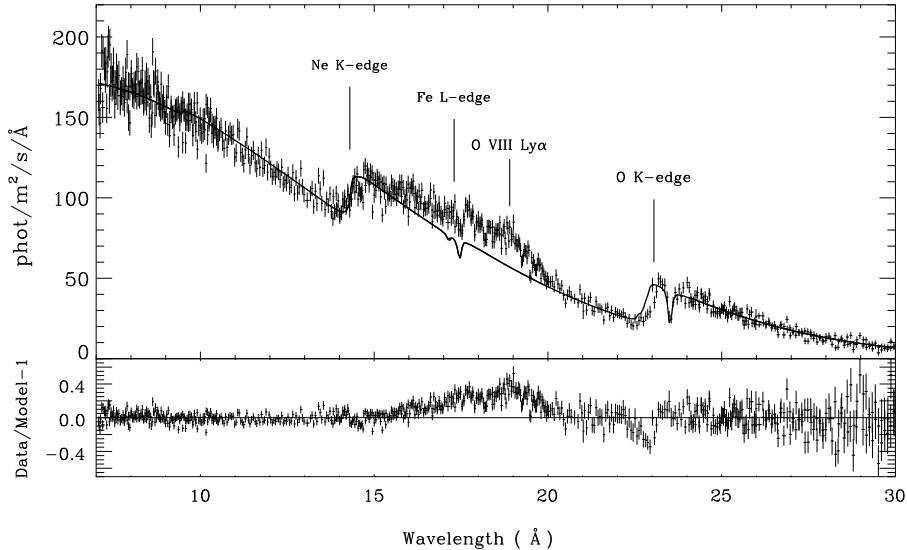}
  \caption{The best-fitting absorbed power-law model overlaid on the RGS spectrum extracted from the second part of the observation of the 4U 0614+091. The residuals show a broad emission feature with a peak at $\sim19$~\AA. The residual at $\sim22.7$~\AA\ is most probably caused by the molecular gas in the interstellar medium, which changes the fine structure of the absorption edge with respect to the atomic gas (see text).} 
\vspace{-3mm}
\end{figure*} 
The data from the MOS1, MOS2 and pn cameras do not agree in the soft part of the spectrum. It was reported that the timing mode calibration of MOS has larger uncertainties than the imaging mode calibration \citep{intZand}. The differences between MOS2 and pn spectra below 1 keV may indicate also uncertainties in the pn timing mode calibration. 
\begin{table}
\begin{center}
\caption{Parameters from the best fit of the model to the MOS2 data. Parameters fixed to the values found in the local fits to the RGS spectra are marked in italics.}
\begin{tabular}{lcc}
\hline
\hline
Parameter  &  \multicolumn{2}{|c|}{MOS2} \\
&model-MOS&model-RGS \\
\hline
$\Gamma$&$2.23\pm0.01$ & $2.220\pm0.006$\\
$N_{H} [10^{21} {\rm cm}^{-2}] $& $3.24\pm0.04$& \textit{3.04}\\
$A_{O}^{*}$ &$\textbf{0.50}\pm0.03$ & \textit{\textbf{1.0}}  \\
$A_{Ne}$ &$\textbf{4.3}\pm0.2$ & \textit{\textbf{3.8}}\\
$A_{Fe}$ & 1.0**& \textit{0.23}\\
$E_{Gauss}$ [keV]&-&\textit{0.67}\\
$\sigma_{Gauss}$ [keV]&-&\textit{0.14}\\
$\chi^{2}_\nu$ (d.o.f.)&1.3 (279)  & 1.13 (282)\\
\hline
\hline
\end{tabular}
\end{center}
{\footnotesize*Note the difference between the abundances obtained from fitting MOS and RGS data separately. \\
**Parameter fixed during fitting.}
\vspace{-3mm}
\end{table}
Therefore we focus on the MOS2 data only. In the case of the MOS1 and pn data we report only the fit in the range $1-10$ keV. The fit gives a slope of the power-law of $2.173\pm0.004$ with $\chi^2_{\nu}=2.6$ for 242 d.o.f. for MOS1 and $2.216\pm0.001$ with $\chi^2_{\nu}=2.1$ for 735 d.o.f. for the pn.
We fit MOS2 in the range $0.4-10$ keV assuming an interstellar medium (ISM) with proto-solar abundances. The best-fit model leaves a large residual in the soft X-ray part of the spectrum (Fig. 1 upper left panel). The possible solution proposed by \citet{juett} is to allow the oxygen and neon abundances, whose absorption edges play a significant role in this region, to vary while fitting. Although the fit is much better, the residuals clearly show that the fit can still be improved (Fig.1 upper right panel). In addition the fit gives an unexpected significant underabundance of oxygen (see Table 1, column model-MOS).
\subsubsection{Constraining abundances from RGS data}
Since the spectral resolution of the MOS2 data is not sufficient to resolve the neon and oxygen edge, we analyse the RGS data. Since the source is variable, we focus on the second part of the observation, which is simultaneous with the MOS2 observation. We use the slope of the power-law determined by fitting the MOS2 data as starting value while fitting RGS data. We do not fix the parameter because of the possible cross-calibration uncertainties. The high-resolution spectra of the source show a prominent Ne K-edge at 14.3~\AA, Fe L-edge at 17.3~\AA\ and O K-edge at 23.05~\AA\ (Fig. 2). We measure the depth of these absorption edges by fitting the spectra locally with the {\sc slab} model \citep{kaastra2002}. This model calculates the transmission of a slab of gas containing individually selected elements with the same atomic absorption cross-sections as used in the neutral absorption model. The model contains the detailed structure of the edges. For this purpose we take the following energy ranges: $21-24$~\AA\ for the O K-edge, $16-18$~\AA\ for the Fe L-edge and $12-16$~\AA\ for the Ne K-edge. For these local fits the power-law slope is fixed to the one found from the overall fit. The results are included in the Table 2. \\ 
The abundance of oxygen in the ISM agrees with the proto-solar abundance of \citet{lodders}, contrary to the results of \citet{juett}. As for the abundance of neon, we find an overabundance of $\sim3.8$ over proto-solar, which supports the results obtained by \citet{paerels}. The amount of iron, that we measure is significantly lower than what we would expect from the ISM with proto-solar abundances. 
\begin{center}
\begin{figure*}  
\vspace{-6mm}
\includegraphics[width=0.72\textwidth]{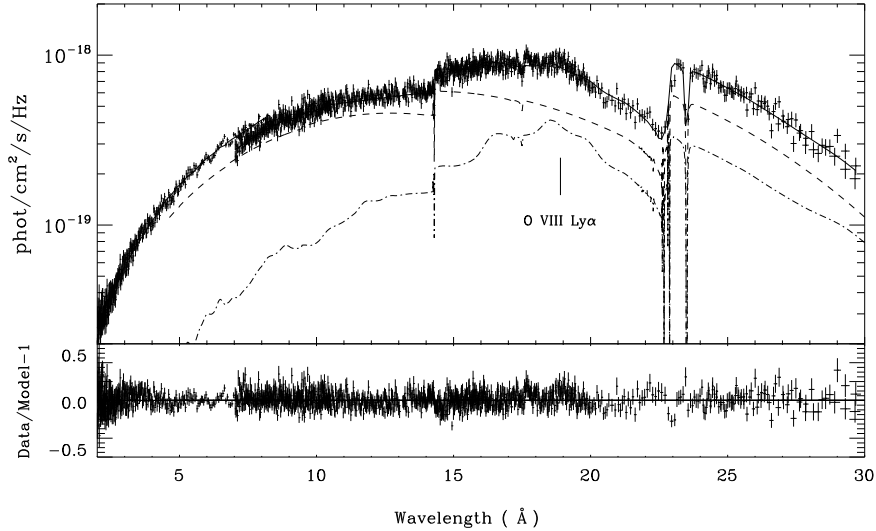}
  \caption{The best-fitting absorbed power-law model with a reflection component overlaid on the RGS and MOS2 spectrum of 4U 0614+091. The power-law and reflection components are additionally plotted in the dashed and dashed-dotted lines respectively.} 
\vspace{-6mm}
\end{figure*} 
\end{center}
\subsubsection{A feature at $\sim$ 19 \AA}
We refit the full RGS spectra with the column densities of oxygen, iron and neon fixed to the values derived from the local fits. The fit gives $\chi^2_{\nu}=2.97$ for 685 d.o.f. and the residuals show structure around 19~\AA\ (0.7 keV). Therefore we refit the RGS spectrum excluding the range $16-21$~\AA. We detect a broad feature with a peak at 19~\AA\ in both parts of the \textit{XMM-Newton} observation (Fig. 2 is made using only the second part of the observation). Modeling the feature with a Gaussian line reveals the center to be at $\sim18.6$~\AA\ with a FWHM of $\sim4$~\AA\ (see Table 2) and equivalent width of  1.1~\AA\ (41 eV), which is calculated by taking the flux of the continuum and Gaussian line in the range $16-21$~\AA. \\ 
We use abundances determined from the second part of the RGS observation to refit the simultaneous MOS2 data with the column densities of each element fixed. The fit reveals an emission feature at around 0.7 keV (Fig. 1 lower left panel). The inclusion of the Gaussian line with the parameters fixed to those found in the RGS data improves the fit to the MOS2 data further with $\chi^2_{\nu}=1.13$ for 282 d.o.f. (Table 1, Fig. 1, lower right panel).
\begin{table}
\begin{center}
\caption{Parameters from the best fit to the RGS data.}
\begin{tabular}{lcc}
\hline
\hline
& \multicolumn{2}{|c|}{RGS}  \\
Parameter&part 1 &part 2\\
\hline
$\Gamma$ &$2.37\pm0.03$ &$2.26\pm0.02$ \\
$N_{\rm H} [10^{21} {\rm cm}^{-2}]$&$3.2\pm0.05$ &$3.04\pm0.04$  \\
$A_{O}$ &$0.93\pm0.06$ &$1.01\pm0.04$ \\
$A_{Ne}$&$4.2\pm0.2$  & $3.8\pm0.15$\\
$A_{Fe}$ &$0.48\pm0.13$&$\textbf{0.23}\pm\textbf{0.07}^{*}$ \\
$\chi^2_{\nu}\slash d.o.f.$ &1.82/519 &2.97/685\\
\hline
&Gaussian profile&\\
\hline
$\lambda_{Gauss}$ [\AA] &$18.64\pm0.17$&$18.56\pm0.10$\\
$\sigma_{Gauss}$ [\AA]&$4.60\pm0.45$&$3.87\pm0.28$\\
$EqW$[\AA]&$1.31\pm0.14$&$1.1\pm0.1$\\
$\chi^2_{\nu}\slash d.o.f.$&0.98/516 &1.38/682\\
\hline
\hline
\end{tabular}
\end{center}
{ \footnotesize *Note low abundance of iron (see Sec. 4). }
\end{table}
\subsection{The broad emission feature: relativistic or Compton scattering broadening ?}
Whereas the Gaussian line on-top of a power-law continuum describes the data, the model lacks a physical interpretation. A plausible interpretation of the emission feature is that of a relativistically broadened emission line. Therefore we fit the Laor profile \citep{laor} to the residuals present in the RGS data. Since the peak of the feature is very close to the energy of the O VIII Ly$\alpha$, we fix the wavelength of the line to 18.97~$\AA$.  We fix the outer radius to 1000$~GM/c^2$. The fit reveals an unexpectedly high inclination of 88 deg (Table 3). With such a high inclination, we expect to observe eclipses in the light curve, which are not present. It could be the case that the inner part of the disc, where the photons are reflected, has a different inclination than the binary system. The inner part of the disc can be warped and twisted if it is strongly irradiated by the neutron star in a non-uniform way \citep{pringle}. The fit gives an inner radius of $\sim14~GM/c^2$ with $\chi_{\nu}^2=1.44$ for 681 d.o.f., however we find also another solution $r_{in}\sim3.5~GM/c^2$ with $\chi_{\nu}^2=1.45$.\\
An additional mechanism broadening the line is Compton scattering. The Compton downscattering can be estimated using the formula $\sigma_{E}/E=E\tau^2/m_{e}c^2$, where $\sigma_{E}$ is Half Width at Half Maximum of the Gaussian line and $E$ is the energy of the line in keV. The calculation gives $\tau\approx9$, which seems unlikely, since it indicates an extremely optically-thick layer of material. This makes it impossible that Compton downscattering alone is responsible for the line broadening. Therefore, the plausible solution points towards broadening by the strong gravitational field and scattering off the ionized material of the disc. 
\subsection{Adding a reflection model}
We fit a relativistically broadened reflection model {\sc reflionx} \citep{ross}, available in {\sc xspec} \citep{arnaud}, which takes into account effects of Compton scattering in the ionized reflecting material. For the relativistic broadening we use the convolution model {\sc kdblur} (based on Laor model) and for the Galactic absorption we use the {\sc TBnew} model (\citeauthor{wilms}, in prep). In order to fit the spectrum with the reflection model we need a significant overabundance of oxygen. In order to mimic the enhancement of oxygen, we decrease the abundance of iron by a factor of 5. Using the described fit-function we find that the $\chi^2$ distribution around the (local) minimum for the inner radius does not allow for an error screening on the best-fit $r_{in}$ values using the current data set. Because of additional complexity of the reflection model, we fix the inner disc radius to that of the innermost stable circular orbit for a non-rotating neutron star (6 $GM/c^2$). The outer disc radius is again fixed to 1000 $GM/c^2$. In the fit we use RGS in the same energy range as before and MOS2 in the range from 1.5 to 10 keV, in order to have good constraints on the power-law slope and the abundances of oxygen, iron and neon. The best fit yields a $\chi_{\nu}^2=1.28$ for 1454 d.o.f. The best-fit inclination is $\sim57$ deg and the best-fit ionization parameter is $\sim222$ erg cm s$^{-1}$ (Table 3), which indicates that indeed some Compton scattering is present as well as relativistic broadening effects. 
\section{Discussion}
Using high-resolution spectra we have shown that a model with an overabundance of neon and underabundance of oxygen cannot fully describe the soft X-ray spectra of 4U 0614+091, contrary to the earlier suggestions \citep[e.g.,][]{juett}. The spectra show an emission line-like feature with a peak at  $\sim19$~\AA. We investigate a possible interpretation of this feature as relativistically broadened emission line of O VIII Ly$\alpha$ and discuss the uncertainties introduced by using only one ISM component (neutral gas). During the refereeing process of this Manuscript the work of \citeauthor{schulz} was submitted, confirming the presence of this broad O VIII emission line using \textit{Chandra} data.\\
The fact that we do not observe the Fe K$\alpha$ line in the broad band spectra indicates that the abundances in the accretion disc differ from solar. Optical spectra of this source show that this is indeed the case as expected from a white dwarf donor star. \citet{nele} find emission lines of oxygen and carbon. We detect also a significant overabundance of neutral neon with respect to the ISM, which probably originates from the source \citep{juett}. This points towards a CO or an ONe white dwarf as the donor star. In both cases, the amount of oxygen in the disc is greatly enhanced with respect to LMXBs with a main-sequence mass donor. The model of \citet{ballant} predicts that with the significant amount of oxygen in the disc the O VIII line becomes the most prominent line in the reflection spectrum. We have shown that using a currently available reflection model it is possible to get a good fit by lowering the abundance of iron. The fit gives a reasonable value of the inclination together with significant ionization of the disc material. 
\begin{table}
\begin{center}
\caption{Parameters from the best fit to the second part of the observation: RGS data fitted with an absorbed power-law and a Laor profile and combined RGS and MOS2 data fitted with an absorbed power-law and a reflection model.}
\begin{tabular}{lccc}
\hline
\hline
&\multicolumn{2}{|c|}{Laor profile}&Reflection model\\
parameter &\multicolumn{2}{|c|}{RGS-part 2}&RGS \& MOS2\\
\hline
$r_{in}$ & $3.5\pm0.4$&$14\pm1$&6.0$^{*}$\\
$q$ &$2.24\pm0.07$&$2.3\pm0.1$& $2.56\pm0.03$\\
$i$ [deg]&$88.0\pm0.1$&$86.8\pm0.2$&$57.1\pm0.5$\\
$\xi$ [erg cm s$^{-1}$]&-&-&$222^{+22}_{-10}$\\
$\chi^2_{\nu}\slash d.o.f.$&1.45/681&1.44/681&1.28/1454\\
\hline
\hline
\end{tabular}
\end{center}
{\footnotesize *Parameter fixed during the fitting.}
\end{table}
\\ In our analysis we took into account only neutral gas ISM components. Since the source spectrum has substantial absorption of the Galactic ISM, we expect other components like dust, molecular gas, warm and hot ionized gas to play a role as well. We find low abundance of iron, which can be explained by the screening effect in the ISM dust grains \citep[see][]{kaastra2009} indicating iron bound in dust in the line of sight. The residual at $22.7-23.2$~\AA\  shows that oxygen may be bound in molecules, which significantly changes the fine structure of the edge compared to the atomic case \citep{pinto}. The narrow features close to the neon edge hint at the presence of ionized neon indicating warm gas, however we cannot constrain the amount owing to the limited signal-to-noise ratio of the data. The hot gas in the outer parts of the Galaxy, where the source is placed, originates from supernova remnants. The radio map of 4U 0614+091 shows no evidence of a  supernova remnant in the line of sight (Miller-Jones, private communication). \\
We attempted to constrain the inner radius of the accretion disc, however we obtained two solutions with comparable $\chi^2$ value: $r_{in}\sim3.5$ and 14$~GM/c^2$. Deeper observations are necessary to measure the inner radius accurately.
\section*{Acknowledgments}
We thank Jelle Kaastra, Elisa Costantini and the referee for useful comments. PGJ acknowledges support from a VIDI grant from the Netherlands Organisation for Scientific Research.

 \bibliographystyle{mn2e}
 \bibliography{mybib1}

%\begin{thebibliography}{99}

%\end{thebibliography}

\end{document}